\documentstyle[12pt]{article}
\title{Comment on ``Neutrino masses and mixing angles
in a predictive theory of fermion masses''}
\author{L.\ Lavoura\thanks{On leave of absence
from Universidade T\'ecnica de Lisboa,
Lisbon, Portugal}\hspace{1mm}
and Jo\~ao P.\ Silva \\
\small Department of Physics, Carnegie-Mellon University, \\
\small Pittsburgh, Pennsylvania 15213, U.S.A.}
\begin{document}
\maketitle
\begin{abstract}
In the extension of the Dimopoulos--Hall--Raby model
of the fermion mass matrices
to the neutrino sector,
there is an entry in the up-quark and neutrino Dirac mass matrices
which can be assumed to arise from the Yukawa coupling of a {\bf 120},
instead of a {\bf 10} or a {\bf 126},
of SO(10).
Although this assumption leads
to an extra undetermined complex parameter in the model,
the resulting lepton mixing matrix exhibits the remarkable feature
that the $ \nu_{\tau} $ does not mix with the other two neutrinos.
Making a reasonable assumption about the extra parameter,
we are able to fit the large-mixing-angle MSW solution
of the solar-neutrino problem,
and we obtain $ m_{\nu_{\tau}} \sim 10 $ eV,
the right mass range to close the Universe.
Other possibilities for explaining the solar-neutrino deficit
are also discussed.
\end{abstract}

\vspace{5mm}

This is a Comment on the paper of Ref.\ \cite{raby}.
In that paper,
the Dimopoulos--Hall--Raby (DHR) \cite{rabymodel}
model of the quark and charged-lepton mass matrices in a SUSY GUT,
which is in fact a revival of the old Georgi--Jarlskog \cite{georgi} scheme,
has been extended,
with the help of some assumptions,
to yield also the form of the neutrino mass matrices,
thus predicting the neutrino mass ratios and the lepton mixing.
In this Comment,
we show that,
if one of the assumptions is relinquished,
a different form of the neutrino Dirac mass matrix can be obtained,
which leads to a very interesting prediction for the lepton mixing,
namely,
that the $ \tau $ neutrino is an eigenstate of mass.

The DHR model
requires a small number of SO(10) Higgs representations
to give mass to the fermions.
It requires one $ {\bf 10}^d $ and one $ {\bf 126}^d $
to give mass to the down-type quarks and to the charged leptons,
and one $ {\bf 126}^{uN} $ and one other representation
to give mass to the up-type quarks and to the neutrinos.
For this other representation,
DHR have used either a $ {\bf 126}^u $
--- obtaining results which agree only marginally
with the MSW solution of the solar-neutrino problem ---
or a $ {\bf 10}^u $
--- obtaining results which do not agree with that solution.
We remark in this Comment that
the quark- and charged-lepton-sector predictions
are the same if one uses instead a $ {\bf 120}^u $ as that representation.
In looking for predictions for the neutrino sector,
inspired by the successes obtained in the other sectors,
all possibilities should be considered.
In this Comment,
we complement the DHR analysis by studying the case with the $ {\bf 120}^u $,
and show that that case leads to dramatically different physics.

The DHR model,
as perfected in Ref.\ \cite{raby},
is appealing for the following reasons.
First of all,
it has great predictive power;
in that feature,
it is matched by other more recent models \cite{giudice,shrock}.
The DHR model can be enforced by means of a simple $ Z_n $ symmetry,
$ n \geq 5 $,
on the Yukawa couplings:
with $ \omega^n = 1 $,
the first $ {\bf 16} $ of fermions transforms with $ \omega $,
the second one with $ \omega^3 $ and the third one with $ \omega^2 $,
while the $ {\bf 10}^d $ and the $ {\bf 126}^{uN} $
transform with $ \omega^{-4} $,
the $ {\bf 126}^d $ with $ \omega^{-6} $,
and the $ {\bf 120}^u $ with $ \omega^{-5} $.
Finally,
the DHR model does not make unjustified assumptions,
like some mass-matrix elements being equal to other mass-matrix elements,
or some phases of mass-matrix elements happening to vanish.
These features make the DHR model very attractive.

We assume that the (2,3) and (3,2) elements
of the up-quark mass matrix arise from the Yukawa couplings
of a $ {\bf 120}^u $.
The mass matrices of the charged fermions,
at the GUT scale,
read
\begin{eqnarray}
M_D & = & \left( \begin{array}{ccc}
0 & F \exp (i \alpha_1) & 0 \\
F \exp (i \alpha_1) & E \exp (i \alpha_2) & 0 \\
0 & 0 & D \exp (i \alpha_3)
\end{array} \right)\, ,
\label{eq:MD}\\
M_E & = & \left( \begin{array}{ccc}
0 & F \exp (i \alpha_1) & 0 \\
F \exp (i \alpha_1) & - 3 E \exp (i \alpha_2) & 0 \\
0 & 0 & D \exp (i \alpha_3)
\end{array} \right)\, ,
\label{eq:ME}\\
M_U & = & \left( \begin{array}{ccc}
0 & C \exp (i \alpha_4) & 0 \\
C \exp (i \alpha_4) & 0 & - B \exp (i \alpha_5) \\
0 & B \exp (i \alpha_5) & A \exp (i \alpha_6)
\end{array} \right)\, .
\label{eq:MU}
\end{eqnarray}
$ A $,
$ B $,
...,
$ E $ and $ F $ are real and positive by definition.
All except one phase can be rotated away,
and therefore the minus sign in $ (M_U)_{23} $
does not affect the analysis \cite{rabymodel} of the DHR model.
That phase is
$ \xi \equiv - \alpha_1 + \alpha_2 + \alpha_4 - 2 \alpha_5 + \alpha_6 - \pi $.
For instance,
we find for the Cabibbo angle the result
\begin{equation}
V_{us} \approx
\sqrt{\frac{m_d}{m_s}} + e^{i \xi} \sqrt{\frac{m_u}{m_c}}\ .
\label{eq:cabibboangle}
\end{equation}
DHR \cite{rabymodel} fitted the Kobayashi--Maskawa matrix in their model,
and found $ \cos \xi = 0.38^{+0.21}_{-0.14} $ at the weak scale.

Our Majorana mass matrix for the right-handed neutrinos
is exactly the same as in the paper commented upon:
\begin{equation}
M_{NN} \propto \left( \begin{array}{ccc}
0 & C \exp (i \alpha_4) & 0 \\
C \exp (i \alpha_4) & 0 & 0 \\
0 & 0 & A \exp (i \alpha_6)
\end{array} \right)\, .
\label{eq:majoranamass}
\end{equation}
The neutrino Dirac mass matrix $ M_{\nu N} $ has two contributions,
one from the $ {\bf 126}^{uN} $,
the other one from the $ {\bf 120}^u $.
The contribution of the $ {\bf 126}^{uN} $ is equal to
$ (-3) $ times its contribution to $ M_U $.
The contribution of the $ {\bf 120}^u $ is unknown,
because in the $ {\bf 120}^u $ there are two doublets
whose vacuum expectation values (VEVs)
contribute to $ M_U $ and to $ M_{\nu N} $,
and while for one of the doublets the contributions to $ M_U $ and to
$ M_{\nu N} $ are equal,
for the other doublet the contribution to $ M_{\nu N} $ is $ (-3) $ times
the contribution to $ M_U $.
Therefore,
\begin{equation}
M_{\nu N} = \left( \begin{array}{ccc}
0 & - 3 C \exp (i \alpha_4) & 0 \\
- 3 C \exp (i \alpha_4) & 0 & - T \exp (i \alpha_7) \\
0 & T \exp (i \alpha_7) & - 3 A \exp (i \alpha_6)
\end{array} \right)\, ,
\label{eq:diracmass}
\end{equation}
with $ T $ real and positive by definition.

{}From Eqs.~\ref{eq:majoranamass} and \ref{eq:diracmass} we find
the effective mass matrix for the light neutrinos:
\begin{eqnarray}
M_{\nu \nu} & = & - M_{\nu N} M_{NN}^{-1} M_{\nu N}^T
\nonumber\\
            & + &
\frac{1}{2} M_{\nu N} M_{NN}^{-1}
\{ M_{NN}^{-1} , M_{\nu N}^T M_{\nu N} \}
M_{NN}^{-1} M_{\nu N}^T + ...\, ,
\label{eq:effectivecorrected}
\end{eqnarray}
where $ \{ X , Y \} $ denotes the anti-commutator
of the matrices $ X $ and $ Y $.
We work in the first-order approximation,
in which
\begin{eqnarray}
M_{\nu \nu} & = &
- M_{\nu N} M_{NN}^{-1} M_{\nu N}^T
\nonumber\\
            & \propto &
\left( \begin{array}{ccc}
0 & (C/A) \exp [i (\alpha_4 - \alpha_6)]) & 0 \\
(C/A) \exp [i (\alpha_4 - \alpha_6)] &
(T^2 / 9 A^2) \exp [2 i (\alpha_7 - \alpha_6)] & 0 \\
0 & 0 & 1
\end{array} \right)\, .
\label{eq:effectivemass}
\end{eqnarray}
This equation,
together with Eq.~\ref{eq:ME},
leads to the main prediction of our scheme:
the $ \tau $ neutrino does not mix with the other two.
The lepton mixing,
in our model,
occurs only between the first two generations.
This is a remarkable feature,
which in general would be difficult to obtain.
Notice however
that this is not true any more if one considers
the higher-order corrections to $ M_{\nu \nu} $,
which are suppressed by extra powers of $ M_{NN} $.
If we calculate the second term in the right-hand-side of
Eq.~\ref{eq:effectivecorrected},
we find that $ M_{\nu \nu} $ indeed includes three-generation mixing,
which is however,
for large $ M_{NN} $,
extremely small.

One of the two ratios of neutrino masses is given by
\begin{equation}
\frac{m_1 m_2}{m_{\nu_{\tau}}^2} =
\frac{m_t m_c m_u}{(m_t - m_c + m_u)^3} \approx
\frac{m_c m_u}{m_t^2}\, .
\label{eq:taumass}
\end{equation}
$ m_1 $ and $ m_2 $ are the masses of the two lighter neutrinos,
which are mixtures of $ \nu_e $ and $ \nu_{\mu} $.
The mass ratio $ m_1 / m_2 $ is a function
of the unknown parameter $ T $:
\begin{equation}
\frac{m_2 - m_1}{\sqrt{m_2 m_1}} =
\frac{1}{9} \left( \frac{T}{B} \right)^2
\frac{(m_t - m_c) (m_t + m_u) (m_c - m_u)}{\sqrt{m_t m_c m_u
(m_t - m_c + m_u)^3}}
\approx \frac{1}{9} \left( \frac{T}{B} \right)^2
\sqrt{\frac{m_c}{m_u}}\, .
\label{eq:massratio}
\end{equation}
Notice that,
if $ T $ vanishes,
$ m_2 $ and $ m_1 $ become equal,
and the mixing between $ \nu_e $ and $ \nu_{\mu} $
becomes maximal \cite{smirnov}.
The lepton mixing matrix $ K $ is
\begin{equation}
K = \left( \begin{array}{ccc}
i (c c^{\prime} + s s^{\prime} e^{i \chi}) &
i (- c s^{\prime} + s c^{\prime} e^{i \chi}) & 0 \\
s c^{\prime} - c s^{\prime} e^{i \chi} &
- s s^{\prime} - c c^{\prime} e^{i \chi} & 0 \\
0 & 0 & 1
\end{array} \right)\, .
\label{eq:matrizbp}
\end{equation}
Here,
we have defined $ \chi \equiv \xi + 2 \alpha_5 - 2 \alpha_7 $.
As $ \alpha_7 $ is unknown,
$ \chi $ is also unknown.
We have also defined
\begin{equation}
c \equiv \sqrt{\frac{m_2}{m_2 + m_1}}\, , \hspace{2mm}
s \equiv \sqrt{\frac{m_1}{m_2 + m_1}}\, , \hspace{2mm}
c^{\prime} \equiv \sqrt{\frac{m_{\mu}}{m_{\mu} + m_e}}\, , \hspace{2mm}
s^{\prime} \equiv \sqrt{\frac{m_e}{m_{\mu} + m_e}}\, .
\label{eq:angulos}
\end{equation}
The factor $ i $ in some matrix elements of
$ K $ in Eq.~\ref{eq:matrizbp}
was inserted in order to obtain positive $ m_1 $.
Notice that there is CP violation in the leptonic sector in our model,
due to the presence of the phase $ \chi $ in the mixing matrix $ K $,
and to the Majorana character of the neutrinos.

The relevant mixing parameter $ \sin^2 (2 \theta_{e \mu}) $
is a function of the unknown phase $ \chi $,
and of the real parameter $ (T/B) $,
which fixes the ratio $ (m_1 / m_2) $.
We have depicted that function in Figure 1.
We have also marked in that figure the values of the mixing parameter
which are favored \cite{langacker}
by the MSW \cite{msw}
explanation of the solar-neutrino problem.
We observe that the large-mixing-angle explanation
of the solar-neutrino depletion
is easily obtained for any value of the phase $ \chi $,
with values of $ (T/B) $ close to 1.
On the other hand,
the small-mixing-angle explanation
can only be obtained for $ \cos \chi > 0.5 $,
and this when the value of $ (T/B) $ is about 3.
This is best seen in Figure 2.

These results justify the additional assumptions that,
in the $ {\bf 120}^u $ of SO(10),
only the doublet in the {\bf 1} of SU(4) \cite{patisalam},
or alternatively only the doublet in the {\bf 15} of SU(4),
acquires a VEV.\footnote{Such assumptions may look unnatural.
However,
the DHR scheme which we are considering makes similar assumptions
from its very beginning anyway.
For instance,
the $ {\bf 126}^{uN} $ is assumed to have VEVs contributing
to $ M_U $ and to $ M_{\nu N} $ on the one hand,
and to $ M_{NN} $ on the other hand,
but no VEV contributing to $ M_D $ and $ M_E $.
Similarly,
the $ {\bf 10}^d $ has a VEV in the $M_D$--$M_E$ direction,
but no VEV in the $M_U$--$M_{\nu N}$ one.
And so on.}
The first assumption gives $ (T/B) = 1 $ and the second assumption
gives $ (T/B) = 3 $.
Both assumptions also fix $ \chi $ to be equal to $ \xi $.
Remember that the phase $ \xi $ can be found from a fit of the CKM matrix,
and one obtains a positive value for $ \cos \xi $.
The assumption that only the doublet in the {\bf 1} of SU(4)
acquires a VEV leads to
the large-mixing-angle solution of the solar-neutrino problem,
provided $ \cos \xi \approx 0 $ at the GUT scale.
The alternate assumption that it is the doublet in the {\bf 15} of SU(4)
which has a VEV is also possible,
leading to the small-mixing-angle solution of the solar-neutrino problem,
for $ \cos \xi \sim 0.6 $ at the GUT scale.

Let us now consider the prediction for the mass of the $ \tau $ neutrino.
This can be obtained from Eq.\ \ref{eq:taumass},
provided both $ (m_1 / m_2) $ and the overall scale of the masses
$ m_1 $ and $ m_2 $ are known.
We take $ (m_1 / m_2) $ from the fit of $ \sin^2 (2 \theta_{e \mu}) $,
and we use $ m_2^2 - m_1^2 \sim 10^{-5} {\rm eV}^2 $,
from the MSW explanation of the solar-neutrino deficit.
We obtain that,
for $ T/B = 1 $,
$ m_{\nu_{\tau}} \sim 10 $ eV,
while for $ T/B = 3 $,
$ m_{\nu_{\tau}} \sim 1.5 $ eV.
In the first case,
$ m_{\nu_{\tau}} $ is in the right range to close the Universe
(also notice that for $ T / B > 1 $,
one obtains $ m_{\nu_{\tau}} $ higher than 10 eV).
This corresponds,
just as in the DHR paper \cite{raby},
to having the largest matrix element of $ M_{NN} $ being of order
$ 10^{14} $ GeV.
This is two orders of magnitude below the unification scale,
and suggests either
the existence of an intermediate breaking scale in the model,
or $ M_{NN} $ being an effective mass matrix
coming from higher-dimensional operators \cite{boss}.

Another possibility to explain the solar-neutrino deficit is
the so-called ``just so'' solution,
of large-wavelength neutrino oscillations.
This solution requires \cite{justso} $ \sin^2 (2 \theta_{e \mu}) > 0.85 $
and $ m_2^2 - m_1^2 \sim 10^{-10} {\rm eV}^2 $.
This can be fitted in our model,
for any value of $ \chi $,
if $ (T/B) < 0.5 $.
This explanation of the solar-neutrino deficit has,
in the context of our model,
the advantage that it leads to a very low $ \nu_{\tau} $ mass,
of order $ 0.1 $ eV.
This gives $ M_{NN} \sim 10^{16} $ GeV,
eliminating the need for an intermediate scale.

In conclusion,
we have suggested a model for the fermion mass matrices in SO(10),
which model can be enforced by means of a discrete symmetry.
In our model the $ \nu_{\tau} $ does not mix with the other two neutrinos.
The atmospheric-neutrino puzzle
then cannot be explained by $\nu_{\mu}$--$\nu_{\tau}$ mixing,
and experiments looking for that mixing should obtain a null result,
contrary to the predictions of the DHR model \cite{raby}.
Contrary to the DHR model,
in our model the large-mixing-angle MSW solution
of the solar-neutrino puzzle can be fitted,
with $ m_{\nu_{\tau}} \sim 10 $ eV,
which closes the Universe.
The small-mixing-angle MSW explanation of the solar-neutrino deficit,
is also possible,
for smaller $ m_{\nu_{\tau}} $.
In both cases,
one must introduce an intermediate scale,
to justify the high value of $ m_{\nu_{\tau}} $.
This is not so if one uses our model to fit instead the ``just so'' solution
of the solar-neutrino problem,
in which case one obtains $ m_{\nu_{\tau}} \sim 0.1 $ eV.

\vspace{2mm}

We thank L.\ Wolfenstein for reading the manuscript.
This work was supported by the United States Department of Energy,
under the contract DE-FG02-91ER-40682.
The work of J.\ P.\ S.\ was partially supported by JNICT (Portugal),
under CI\^ENCIA grant BD/374/90-RM.

\vspace{5mm}

%

\vspace{10mm}

{\large \bf FIGURE CAPTIONS}

\vspace{5mm}

Figure 1: Graph of $ \sin^2 (2 \theta_{e \mu}) $ as a function of $ (T/B) $,
for various values of $ \cos \chi $.
The lowest curve,
which goes out of the figure down to $ \sin^2 ( 2 \theta_{e \mu} ) = 0 $,
corresponds to $ \cos \chi = 1 $.
The other four curves,
from the lowest one to the highest one,
correspond to $ \cos \chi $ equal to 0.5,
0,
-0.5 and -1,
respectively.
Also shown are the windows of values of $ \sin^2 ( 2 \theta_{e \mu} ) $
suggested by the MSW explanation of the solar-neutrino deficit:
the large-angle solution is marked by dotted lines,
and the small-angle one by dashed lines.

\vspace{5mm}

Figure 2: Regions of the $\cos \chi$--$(T/B)$ plane
for which one obtains $ \sin^2 ( 2 \theta_{e \mu} ) $
in agreement with the MSW solution of the solar-neutrino problem.
The left-most region corresponds to the large-angle solution,
and covers all values of $ \chi $.
The other region corresponds to the small-angle solution,
and covers only values of $ \chi $ with $ \cos \chi > 0.5 $.
\end{document}